\documentclass[doublecol,linenumbers]{epl2} 

\title{Explosive oscillation death in coupled Stuart-Landau
oscillators}
\shorttitle{Explosive oscillation death} 

\author{Hongjie Bi\inst{1} \and Xin Hu\inst{1} \and Xiyun Zhang\inst{1} \and Yong Zou\inst{1,2} \and Zonghua Liu\inst{1,2} \and Shuguang Guan\inst{1,2}}
\shortauthor{Hongjie Bi \etal}

\institute{
  \inst{1} Department of Physics, East China Normal University, Shanghai, 200241, China\\
  \inst{2} State Key Laboratory of Theoretical Physics, Institute of Theoretical Physics, Chinese Academy of Sciences, Beijing 100190, China }


\abstract{
Recently, the explosive phase transitions, such as explosive percolation and explosive synchronization, have attracted extensive research interest. So far, most existing works investigate Kuramoto-type models, where only phase variables are involved. Here, we report the occurrence of explosive oscillation quenching in a system of coupled Stuart-Landau oscillators that incorporates both phase and amplitude dynamics. We observe three typical scenarios with distinct microscopic mechanism of occurrence,
i.e., ordinary, hierarchical, and cluster explosive oscillation
death, corresponding to different frequency distributions of oscillators, respectively. We carry out theoretical analyses and obtain the backward transition point, which is shown to be independent of the specific frequency distributions. Numerical results are consistent with the theoretical prediction.
}

\begin{document}

\maketitle

\section{Introduction }
Recently, two remarkable phenomena, i.e., explosive percolation (EP) and explosive synchronization (ES), have attracted great attention in the area of nonlinear dynamics and complex networks.

EP refers to the sudden formation of an inter-connected giant
cluster in a growing network, whose size is comparable with that of the whole network \cite{Achlioptas2009,Ziff2009,Radicchi2009,Araujo2010,Nagler2011,Costa2010,Riordan2011,Grassberger2011,Cho2013}. It is shown that EP could occur in a network under Achlioptas-like growth rules, which generally enhance forming small clusters and suppress forming giant clusters before transition \cite{Achlioptas2009}. Such processes are essentially different from the traditional percolation transition in Erd{\"o}s-R{\'e}nyi network, where the size of the giant cluster  grows gradually and continuously after the transition.

ES is an abrupt, irreversible synchronization first reported in a generalized Kuramoto model on a scale-free network when the natural frequency and degree of the $j$th oscillator in the network are correlated, i.e., $\omega_j=k_j$ \cite{Gomez-Gardenes2011}.
This result is drastically different from the continuous synchronization transition found in the Kuramoto model for decades, and thus has stimulated intensive research along this line \cite{Leyva2012,Peron2012,Leyva2013,Liu2013,Ji2013,Zhang2013,Zou2014,Zhang2014}.

So far, ES is only observed in phase synchronization, either in the Kuramoto-type model or in the coupled Rossler system. One important question is: could explosive behaviors occur for the amplitude dynamics in networked oscillators? Motivated by this idea, we investigate a model of coupled Stuart-Landau (SL) oscillators in this paper. By introducing similar coupling scheme as in Ref. \cite{Zhang2013}, we found that explosive oscillation death (EOD) could happen in the model. Specifically, we have numerically identified three typical types of EOD, namely, ordinary EOD (OEOD), hierarchical EOD (HEOD), and cluster EOD (CEOD), corresponding to different distributions of natural frequency of SL oscillators. Interestingly, for all three cases, it is found that the backward phase transitions are independent of the specific \revision{frequency distributions (FDs)}. We further apply a self-consistency analysis, which proves the universal property of the backward phase transitions in the model.


\section{The dynamical model}

Amplitude death (AD) and oscillation death (OD) both \revision{refer} to the complete suppression of oscillations in coupled systems , i.e.,
converting from oscillating states, such as limit cycles or chaos, into fixed points \cite{Saxena2012,Koseska2013}.
In AD all oscillators finally settle on the same fixed point, while in OD they settle on different fixed points.
Since its first discovery in an array of coupled oscillators \cite{Bar-Eli1985}, AD and OD have stimulated continuous research in both theory \cite{Matthews1991,Aronson1990,RamanaReddy1998,Atay2003,Lee2013}
and experiments \cite{RamanaReddy2000,Herrero2000} in the past two decades.
For networked oscillators, it is important to investigate \revision{how transition to AD or OD occurs}, i.e., does it occur suddenly for all oscillators, or develop gradually in the system? This issue has been addressed by several previous works. In Refs. \cite{Yang2007,Ullner2007,Liu2009}, it is found that with the increase of the coupling strength, partial oscillators in the network first undergo AD. Then more and more oscillators convert into AD state as the coupling strength further increases until all oscillators become fixed points finally. In one word,
AD/OD occurs successively in such networked systems. Moreover, in Ref. \cite{Prasad2010}, although AD is shown to occur suddenly for all oscillators in the network, no hysteresis has been reported there.
In this work, our primary concern is whether AD/OD could occur in an explosive way accompanied by a hysteresis in certain coupled oscillators system. To this end, we study a dynamical model
of networked SL oscillators with frequency-weighted coupling \cite{Zhang2013}, i.e.,
\begin{equation}\label{model}
\dot{z}_j(t)=(a+iw_j-|z_j|^{2})z_j(t)+
\frac{K|\omega_j|}{N}\sum_{n=1}^{N}[z_n(t)-z_j(t)].
\end{equation}  ¡¡
Here $j=1,2,\cdots,N$ is the index of oscillators. $z_j(t)=x_j(t)+iy_j(t)$
is the complex amplitude of the $j$th oscillator at time $t$, and the dot represents the time derivative. $a$ is a control parameter for individual SL oscillator, i.e., the dynamics settles on a limit cycle if $a>0$, and on fixed point if $a<0$.
$\omega_j$ is the natural frequency
of the $j$th oscillator, and $K$ the uniform coupling strength. The most important characteristic of this model is that the effective coupling for oscillator $j$ is proportional to its natural frequency $\omega_j$. Therefore, the effective couplings in Eq. (\ref{model}) are heterogeneous rather than homogeneous as in most previous models.

In this work, we consider several typical FDs as listed in Table \ref{Table1}.
Without losing generality, we set $a=1$ and only consider \revision {the situation of global coupling.}
Throughout this paper, numerical integration is carried out by the fourth-order Runge-Kutta method with time step 0.01. The initial phases of the limit cycles are random, i.e, oscillators are uniformly distributed on the unit circle in complex plane at the beginning.

\begin{table*}[htbp]
\begin{center}
\begin{tabular}{|c|c|c|c|c|}
\hline  FD & Formula & Parameter & $K_f$ & $K_b$
\tabularnewline \hline \hline
\raisebox{-1.7ex}[0pt]{Triangle} & $g(\omega) = (\pi\Delta-|\omega|)/(\pi\Delta)^2$, & \raisebox{-1.7ex}[0pt]{$\Delta=0.1$} & \raisebox{-1.7ex}[0pt]{$K_f\approx2.30$} & \raisebox{-1.7ex}[0pt]{$K_b\approx2.02$} \\
        ~ & $|\omega|<\pi\Delta$, 0 otherwise & ~&~&
\tabularnewline \hline
Lorentzian & $g(\omega) = \frac{\Delta}{\pi (\omega ^2 + \Delta ^2)}$   &  $\Delta=0.02$ & $K_f\approx3.32$ & $K_b\approx2.02$
\tabularnewline \hline
 \raisebox{-1.7ex}[0pt]{Uniform} & $g(\omega) = 1/(2\pi\Delta)$, & \raisebox{-1.7ex}[0pt]{$\Delta=0.1$} & $K^1_f\approx1.64$& $K^1_b\approx2.02$ \\
  ~      & $|\omega|<\pi\Delta$, 0 otherwise  & ~& $K^2_f\approx2.02$ & $K^2_b\approx1.56$
\tabularnewline \hline
\end{tabular}\end{center}
\caption{Summary of the three typical FDs with parameters investigated in this paper, and the critical points for both forward ($K_f$) and backward ($K_b$) transitions.} \label{Table1}
\end{table*}

\section{The numerical results}

To characterize the collective behaviors of the coupled SL oscillators, two order parameters can be defined as:
$Re^{i\psi}=\sum_{j=1}^{N}z_j(t)/N$,
and
$R_{\theta}e^{i\phi}=\sum_{j=1}^{N}e^{i\theta_j}/N$.
Here, $\theta_j$ represents the phase of the $j$th oscillator.
Order parameter $R$  ($0\le R \le1$ due to $a=1$ in this study) characterizes the coherence of the complete dynamics, including both amplitude and phase.  Order parameter $R_{\theta}$ ($0\le R \le 1$) only characterizes the phase coherence of the system, which does not involve any information of amplitude.

\begin{figure}[!htpb]
\begin{center}
\includegraphics[width=0.48\textwidth]{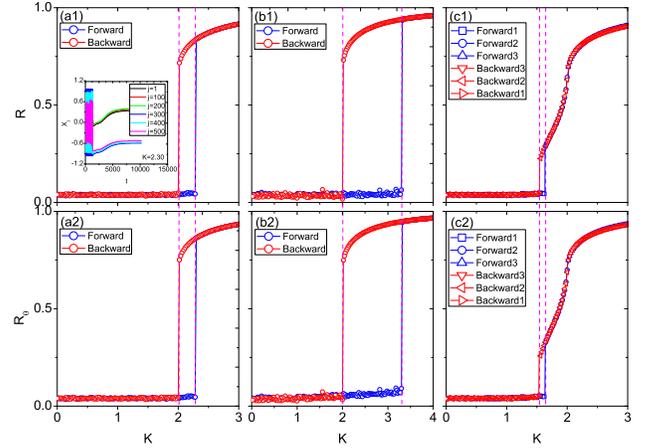}
\caption{(color online)
Characterizing the EOD by order parameters $R$ and $R_\theta$ for three typical FDs, i.e., triangle (the left column), Lorentzian (the middle column), and uniform (the right column), respectively.
System size $N=500$. In both forward and backward transitions,
the dynamical equations are integrated in an adiabatic way, where $K$ is increased at a step of 0.02 and the final state for a prior $K$ is used as the initial state for the next $K$. For each $K$, the order parameters are averaged in a time window after the transient stage. Such numerical schemes are adopted throughout this paper.} \label{fig1}
\end{center}
\end{figure}

We now report the main numerical findings. Generally,
we observe EOD in Eq. (\ref{model}), which, to our knowledge, is the first explosive phase transition involving the behavior of amplitude in a system of coupled oscillators.
As listed in Table \ref{Table1}, we consider three typical FDs in Eq. (\ref{model}).
In Fig. \ref{fig1}, we plot the order parameters $R$ and $R_\theta$ VS the coupling strength $K$. For both $R$ and $R_\theta$, we observe a sudden, discontinuous jump when the coupling strength $K$ exceeds the critical value, i.e., the forward transition point. After the transition, the order parameters are significantly greater than 0, indicating that the system goes from the incoherent state into the (partially) coherent one.
Inversely, when the system starts from a coherent state initially, as the coupling strength decreases, the order parameters also experience an abrupt fall, but at a different coupling strength from the forward transition point. In this way, a hysteresis loop, which is a typical characteristic in the first-order phase transitions, can be defined by the backward and forward transition points.
It should be pointed out that, at this stage,  we cannot identify whether it is OD or AD only based on the results in Fig. \ref{fig1}. Later, we will provide more evidences to confirm that it is OD rather than AD that occurs in this system.

\begin{figure}[!htpb]
\begin{center}
\includegraphics[width=0.48\textwidth]{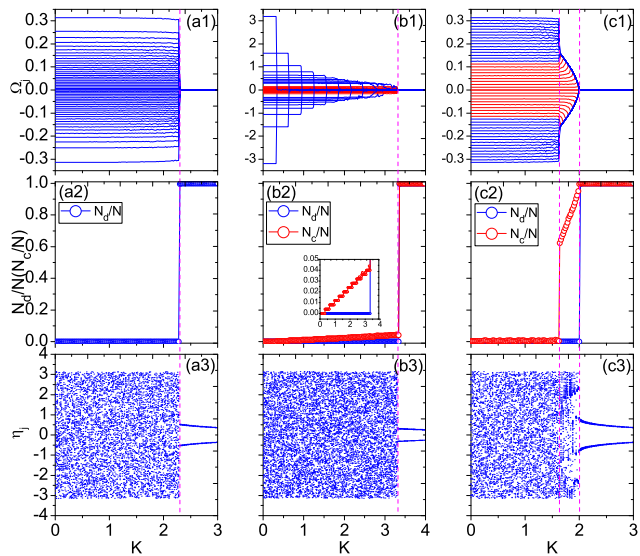}
\caption{(color online) Classifying EOD by three measures: the effective frequencies $\Omega_j$; $N_d/N$ ($N_c/N$);
and the relative phases $\eta_j$ (see the definitions in the text).
For better visualization, we have applied downsampling in this figure when necessary. All parameters are the same as in Fig. \ref{fig1}. In our study, we have also carefully examined the process for the backward transition (not shown). Except for the transition points, the physical pictures are qualitatively the same.}\label{fig2}
\end{center}
\end{figure}

Interestingly, we further reveal that the phenomena of EOD shown in Fig. \ref{fig1} actually correspond to three qualitatively different situations, namely, the OEOD, the HEOD, and the CEOD.
To characterize them, we define several quantitative measures. The first is the effective frequency for each oscillator: $\Omega_j=\frac{1}{T}\int_{t}^{t+T}\dot{\theta_j}(\tau)d\tau$ with $T \gg 1$, which basically is the averaged instantaneous frequency. The second is the relative phase with respect to the average phase:  $\eta_j=\theta_j-\phi$, where $\phi$ is defined as in the order parameter $R_\theta$. \revision{The third are two ratios
$N_d/N$  and $N_c/N$, where $N_d$ and $N_c$ are the numbers of oscillators that are in the OD state and the clustering state, respectively.}
In Fig. \ref{fig2}, we plot these measures for the forward transition processes.
In the following, we describe these three types of EOD in detail.

\begin{figure}[!htpb]
\begin{center}
\includegraphics[width=0.48\textwidth]{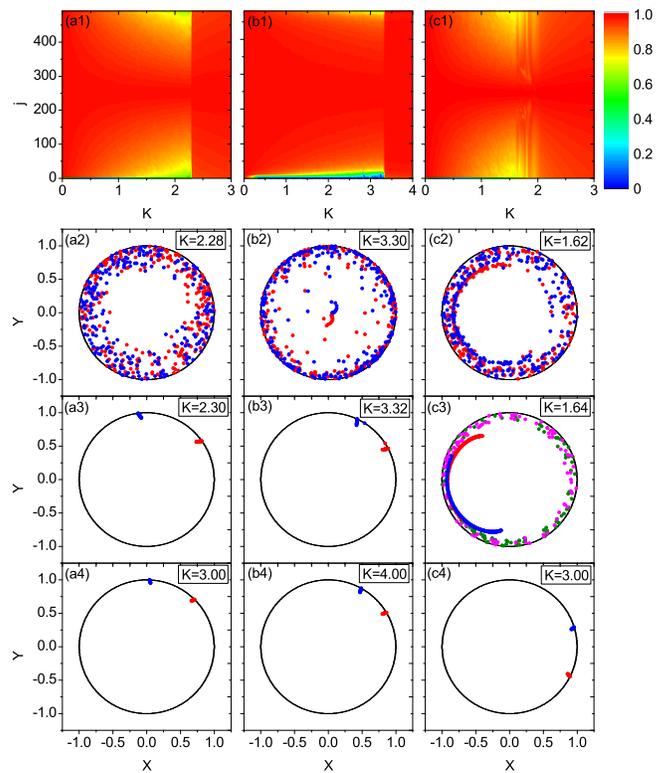}
\caption{(color online)
Characterizing EOD in phase space, corresponding to three FDs, i.e., triangle (the left column), Lorentzian (the middle column), and uniform (the right column), respectively.
In the first row, the amplitudes $|z_j|$, averaged for a long time after transient, are plotted VS $K$. The rest are the snapshots of system states corresponding to different $K$.
The second and the third rows correspond to the states just before and after the first transitions, respectively.
In a2, b2, c2, and c3, red/pinkish-red (blue/green)  dots correspond to oscillators rotating clockwise (counterclockwise).
In a3, b3, a4-c4, red (blue) squares indicate the fixed points converted from oscillators with positive (negative) natural frequencies. The unit circle is plotted to guide the eyes.
All parameters are the same as in Fig. \ref{fig1}. }  \label{fig3}
\end{center}
\end{figure}

\begin{enumerate}

\item {\bf The OEOD}. This corresponds to the triangle FD,  and is basically the same as the situation of ES observed in previous works \cite{Gomez-Gardenes2011,Leyva2012,Peron2012,Zhang2013}. As shown in Figs. \ref{fig2}(a1) and \ref{fig2}(a2), with the increase of $K$, oscillators oscillate almost according to their natural frequencies. This situation continues until $K$ arrives at the forward transition point, where all the effective frequencies of oscillators suddenly and simultaneously collapse to 0. Note that 0 is the averaged natural frequency due to the symmetry of FDs in this study. This transition point characterizes the occurrence of EOD in the system. Moreover, as shown in Fig. \ref{fig2}(a3), after the OD, the oscillators settle on different fixed points in phase space that form two clusters.

\item {\bf The HEOD}. This corresponds to the Lorentzian FD. As shown in Figs. \ref{fig2}(b1) and \ref{fig2}(b2), with the increase of $K$, the effective frequencies of some oscillators collapse hierarchically to small values that are very close to 0.
    This process starts from the largest frequencies and monotonically moves to smaller frequencies.
    A careful examination of the dynamics of individual oscillator reveals that those oscillators whose effective frequencies have collapsed actually form two small clusters, rotating very slowly near the origin on complex plane (Fig. \ref{fig3}(b2)).
    However, since this collapse of effective frequencies before the forward transition point only involves minority oscillators in the system (Fig. \ref{fig2}(b2)), their contributions to the order parameters can be ignored. Therefore, the order parameters have no significant growth at this stage, as shown in Figs. \ref{fig1}(b1) and \ref{fig1}(b2).
    This situation remains until $K$ arrives at the forward transition point, where the effective frequencies of  all oscillators suddenly collapse to 0. Since all oscillators simultaneously become OD (Fig. \ref{fig2}(b2)), the order parameter behaves in an explosive way, as shown in Figs. \ref{fig1}(b1) and \ref{fig1}(b2).

\item {\bf The CEOD}. This corresponds to the uniform FD.
 As shown in Figs. \ref{fig2}(c1) and \ref{fig2}(c2), with the increase of $K$, all oscillators oscillate almost according to their natural frequencies before the first transition point ($K^1_f\approx1.64$). Then when $K$ arrives at this point, those oscillators with relatively large natural frequencies suddenly become locked and form two clusters. Inside each cluster, all oscillators have the same effective frequency, and they behave like two (giant) oscillators, rotating clockwise and counterclockwise, respectively. Since the number of oscillators involved in this clustering is comparable to the system size (Fig. \ref{fig2}(c2)), the order parameters jump explosively, as shown in Figs. \ref{fig1}(c1) and \ref{fig1}(c2). After the first transition, the system is in a partially coherent state, where clusters coexist with drifting oscillators.
 When $K$ is increased further, the two clusters rotate slower and slower. Meanwhile, they gradually absorb the drifting oscillators until finally the effective frequencies of the two giant clusters completely approach 0 at the second critical point ($K^2_f\approx2.02$), where OD occurs. So in this case, the system undergoes two transitions: one is explosive (at $K^1_f\approx1.64$); and the other is continuous (at $K^2_f\approx2.02$). \revision {The transition involves three distinct dynamical states, i.e., the incoherent state, the partially coherent state, and the OD state.} We emphasize that from Figs. \ref{fig1}(c1) and \ref{fig1}(c2), the order parameters actually change continuously when OD occurs at the second critical point. However, since there is an explosive transition in the way toward OD in this case, we still term the whole transition process as CEOD.
\end{enumerate}

In Fig. \ref{fig2}, we have characterized the observed EOD from the perspective of phase, ignoring  the amplitudes of oscillators.
To completely illustrate the physical picture in the above three EOD processes, we now further characterize them in phase space, as shown in Fig. \ref{fig3}. Note that in our numerical simulations, initially all oscillators are randomly distributed on the unit circle in the complex plane.

In the case of OEOD, with the increase of $K$, oscillators gradually move inside the unit circle, as shown in Figs. \ref{fig3}(a1) and \ref{fig3}(a2). However, according to Fig. \ref{fig2}(a1), these oscillators still rotate almost according to their natural frequencies though their amplitudes have been suppressed. Then when $K$ arrives at the forward critical point, all oscillations suddenly cease and OD occurs, as shown in Figs. \ref{fig3}(a2) and \ref{fig3}(a3).  After OD, two compact clusters consisting of different fixed points are formed. \revision{Oscillators inside each cluster may slightly change positions when $K$ further increases. However, oscillators cannot redistribute between two branches  \cite{Koseska2010}.}

In the case of HEOD,
from Fig. \ref{fig2}(b1), we know that the effective frequencies of oscillators hierarchically collapse (near 0), starting from the largest frequency to the smaller ones successively. As shown in Fig. \ref{fig3}(b2), we found that those oscillators form two small clusters near the origin. In the mean time, many oscillators move inside the unit circle, i.e., their amplitudes are suppressed.
Then when $K$ exceeds the forward critical point, all oscillations in the system suddenly stop, indicating the occurrence of OD. Similarly, two compact clusters of fixed points form after OD.

In the case of CEOD, with the increase of $K$, oscillators gradually move inside the unit circle, i.e., the oscillations are significantly suppressed. Even so, they still rotate almost according to their natural frequencies.
Then when $K$ exceeds the first critical point ($K^1_f\approx1.64$), some oscillators, i.e., those with relatively large frequencies, suddenly become frequency-locked and form two clusters. After that, the system enters the partially coherent state, as shown in  Fig. \ref{fig3}(c3).
Remarkably, it is clustering rather than OD that occurs at this point, which
is essentially different from the two cases discussed above.
When $K$ further increases, more and more drifting oscillators are recruited into the coherent clusters.
Meanwhile, the rotations of the two clusters become slower and slower.
Finally, when $K$ exceeds the second critical point ($K^2_f\approx2.02$),  OD occurs, as shown in Fig. \ref{fig3}(c4).

We emphasize that,
because oscillators go to different fixed points in the quenched state, \revision{as shown in the inset of Fig. 1(a1),}  it is OD rather than AD that occurs in this system.
Moreover, in all three cases,
with the further increase of $K$ after
OD, the formed two clusters will become more compact and more approach the unit circle, as shown in Fig. \ref{fig3}.

\section{The theoretical analysis}

In the \revision{previous} section, three typical cases of EOD in networked SL oscillators have been characterized. In all three cases,
the hysteresis loops are observed in phase diagrams. Particularly, it is founded that a backward transition always occurs at $K_b\approx2.02$ (Fig. \ref{fig1} and Table \ref{Table1}), despite of different FDs used.
Theoretically, it is desirable to predict both the forward and backward transition points for the EOD. However, the mathematical treatment turns out to be difficult even for Kuramoto model, which only involves phase variables \cite{Peron2012,Zou2014}. In the following, we apply a self-consistent analysis to obtain the backward transition point and prove that it is independent of specific FDs.

To facilitate the analysis, we consider Eq. (\ref{model}) in polar coordinates, i.e.,
\begin{eqnarray}\label{model-polar}
\dot{r}_j&=&(1-r^2_j-K|\omega_j|)r_j
+\frac{K|\omega_j|}{N}
\sum^N_{n=1}r_n\cos(\theta_n-\theta_j),\nonumber \\ \dot{\theta}_j&=&\omega_j+\frac{K|\omega_j|}{N}
\sum^N_{n=1}\frac{r_n}{r_j}\sin(\theta_n-\theta_j).
\end{eqnarray}
Using the definition of order parameter, the above equation can be written in the mean-field form:
\begin{eqnarray}\label{model-polar2}
\dot{r}_j &=& (1-r^2_j-K|\omega_j|)r_j
+KR|\omega_j|\cos(\psi-\theta_j), \nonumber \\
\dot{\theta}_j &=& \omega_j
+\frac{KR|\omega_j|}{r_j}\sin(\psi-\theta_j).
\end{eqnarray}
We set a reference rotating frame, i.e., $\psi(t)=\psi(0)+\left< \Omega \right> t$, where  $\left<\Omega\right>$ is the average effective frequency of oscillators in the system.
For a symmetric distribution of $g(\omega)$, we generally have $\left<\Omega \right>=0$. Defining $\eta_j\equiv\theta_j-\psi$, Eq. (\ref{model-polar2}) becomes:
\begin{eqnarray}\label{model-polar3}
\dot{r}_j &=&(1-r^{2}_j-K|\omega_j|)r_j
+KR|\omega_j|\cos\eta_j,\nonumber \\
\dot{\eta}_j &=&\omega_j
-\frac{KR|\omega_j|}{r_j}\sin\eta_j.
\end{eqnarray}
When all oscillators are phase-locked, we have $\dot{\eta}_j=0$, then the stationary state for $\eta_j$ can be solved as:
\begin{equation}\label{eta}
 \eta^*_{j}=\left\{\begin{array}{ll}
 \eta^*_+=\arcsin(\frac{r_j}{KR}), & \, \omega_j>0, \\
 \eta^*_-= -\arcsin(\frac{r_j}{KR}),& \, \omega_j<0.
            \end{array} \right.
\end{equation}
Eq. (\ref{eta}) implies that in the phase-locked state, the oscillators in the system will evenly split into two clusters
which are symmetric with respect to the real axis in complex plane. If the coupling strength  $K$ further increases, they both approach the real axis gradually. Only in the limit case of $K\to\infty$, the two clusters coincide into one.
This result has been numerically verified as shown in Fig. \ref{fig3}.

Noticing that the system will form two very compact clusters in phase space when $K$ is large enough, we can approximately regard  the whole system as two oscillators at the centroids of these two clusters. They are located at $(r,\eta_+)$ and $(r,\eta_-)$, \revision{rotating} with frequency $\omega>0$ and $-\omega<0$, respectively. Then Eqs. (\ref{model-polar}) becomes:
\begin{eqnarray}\label{two-oscillators }
\dot{r}&=& (1-r^2-K\omega)r+K\omega r[1+\cos(\eta_+-\eta_-)]/2, \nonumber \\ \dot{\eta}_+&=&\omega+K\omega\sin(\eta_--\eta_+)/2, \nonumber \\
\dot{\eta}_-&=&-\omega+K\omega\sin(\eta_+-\eta_-)/2.
\end{eqnarray}
Set $\Theta=\eta_+-\eta_-$, we obtain:
\begin{eqnarray}\label{two-oscillators-2 }
\dot{r}&=& (1-r^2-K\omega)r+K\omega r[1+\cos\Theta]/2, \nonumber \\ \dot{\Theta}&=&2\omega-K\omega\sin\Theta.
\end{eqnarray}
From above equations, $\Theta$ has steady solution $\Theta^*$
only when $K \ge 2$, i.e.,
\begin{equation}\label{Theta-solution}
\sin\Theta^*=2/K,
\end{equation}
or $\Theta_1^*=\arcsin(2/K)$ \revision{and} $\Theta_2^*=\pi-\arcsin(2/K)$. Further linear analysis shows that the former steady solution is stable while the latter is unstable. For the amplitude equation, the steady solution satisfies the following equation:
\begin{equation}\label{r-solution }
(1-{r^*}^2-K\omega)r^*+K\omega r^*[1+\cos\Theta^*]/2=0.
\end{equation}
When $K\to \infty$, $\Theta^*\to 0$ according to Eq. (\ref{Theta-solution}), thus $\cos\Theta^*\to 1$, and the steady solution of $r$ should obey:
\begin{equation}\label{r-solution-2 }
1-{r^*}^2=0,
\end{equation}
which gives $r^*=1$ for the steady states.

\begin{figure}[!htpb]
\begin{center}
\includegraphics[width=0.38\textwidth]{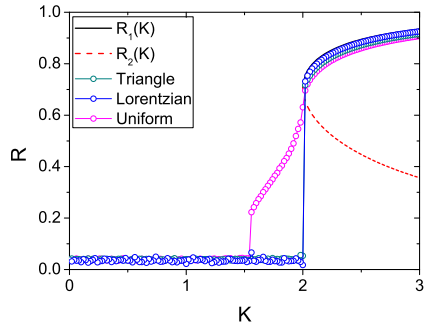}
\caption{(color online)
Comparing the theoretical prediction of backward transition process with numerical simulations. \revision{They agree with each other well, showing that the backward transition is universal, i.e., independent of specific FDs.} } \label{fig4}
\end{center}
\end{figure}

Now, let us solve the order parameter $R$ for the backward transition process. According to definition,
\begin{equation}\label{order parameter-3}
R=Re(\frac{1}{N}\sum_{j=1}^Nr_je^{i\eta_j})
=\frac{r^*}{2}(\cos\eta^*_+ + \cos\eta^*_-).
\end{equation}
Substituting Eq. (\ref{eta}) and $r^*=1$ into Eq. (\ref{order parameter-3}), we get the analytical form of the order parameter as:
\begin{eqnarray}\label{order parameter-4}
R_1(K)&=&\frac{\sqrt{2}}{2}\sqrt{1 + \sqrt{1-4/K^2}},\\
R_2(K)&=&\frac{\sqrt{2}}{2}\sqrt{1 - \sqrt{1-4/K^2}}.
\end{eqnarray}
These two branches corresponds to the stable  and unstable
solutions  of Eq. (\ref{Theta-solution}), respectively.
Since $R$ must be real, Eq. (\ref{order parameter-4}) implies that the (fully) coherent state in the system only exists for $K \geq 2$, i.e., the OD state only exists when $K \geq 2$.  If one continuously decreases  $K$ from $\infty$, the stable and  unstable OD states will gradually approach each other. Finally they collide and  disappear via saddle-node bifurcation at $K^t_b=2$ (the superscript t means theory), which just marks the first backward transition point in the system. Numerical results have verified that in all three cases of EOD, the backward transition occurs at $K\approx2.02$, independent of the \revision{forms} of specific FDs. In Fig. \ref{fig4}, we compare the analytical curves of order parameter with numerical results. It is shown that the backward transition processes with different FDs approximately follow a universal curve. The numerical and
theoretical results are consistent with each other quite well.

\section{Conclusion}

In this paper, we reported the phenomenon of EOD in a system of coupled SL oscillators. By extensive numerical experiments, we classified three typical types of EOD, i.e., OEOD, HEOD, and CEOD, corresponding to triangle, Lorentzian, and uniform FDs, respectively.
In the case of OEOD, all oscillations suddenly cease at a critical point. In the case of HEOD, a few oscillators with large frequencies successively form two \revision{clusters} first, then EOD suddenly occurs for the whole system. In the case of CEOD, there are two transitions towards OD, i.e., first two giant clusters are formed in an explosive way; then they gradually absorb drifting oscillators until all oscillations are completely suppressed finally.
We provided a self-consistent analysis, which enables us to predict a universal backward transition point, and prove that it is independent of the specific FDs. The theoretically results are supported by the numerical observations.

The present work demonstrated that, apart from Kuramoto model,  explosive transition might occur  in certain dynamical system involving amplitudes. Therefore, it broadens our view of explosive phase transition, and helps us better understand the collective behaviors of networked systems. \revision{In our simulations, we have also observed that explosive OD will not happen in the present model if the network topology is scale-free. This raises an important question, i.e., the interplay between network topology and dynamics in the process of OD/AD. In addition, could other coupling schemes also lead to explosive OD? Is it possible to observe explosive AD, and could such AD convert into OD  \cite{Koseska2013b}?   Certainly, these questions deserve further investigation in the future.}

\acknowledgments

This work is supported by the NSFC grants
Nos. 11075056, 11375066, 11305062 and 11135001; the Innovation Program of Shanghai Municipal Education Commission grant No. 12ZZ043; and the Open Project Program of State Key Laboratory of Theoretical Physics, Institute of Theoretical Physics, Chinese Academy of Sciences, China (No. Y4KF151CJ1).

\end{document}